\documentclass{IEEEcsmag}

\usepackage[colorlinks,urlcolor=blue,linkcolor=blue,citecolor=blue]{hyperref}
\expandafter\def\expandafter\UrlBreaks\expandafter{\UrlBreaks\do\/\do\*\do\-\do\~\do\'\do\"\do\-}
\usepackage{upmath,color}
\usepackage{xspace}
\usepackage{enumerate}
\usepackage{microtype}
\usepackage{comment}
\usepackage{booktabs}
\usepackage[table]{xcolor}
\usepackage{makecell}
\usepackage{tabularx}
\usepackage{graphicx}
\usepackage{wrapfig}
\usepackage{multirow}
\jvol{XX}
\jnum{XX}
\paper{8}
\jmonth{Month}
\jname{IEEE Software}
\jtitle{An Evaluation of Role-Based Multi-Agent Code
Generation on Repository-Scale Problems}
\pubyear{2026}

\newcommand{\review}[2]{#2}

\newcommand{\vale}[1]{\colorbox{yellow}{\textbf{VALE:} #1}\xspace}

\usepackage{eso-pic}
\usepackage{hyperref}

\AddToShipoutPictureFG*{%
\AtPageLowerLeft{%
\raisebox{-2.3cm}{%
\hspace*{5.0cm}%
\parbox{\dimexpr\paperwidth-5.0cm\relax}{
This article has been accepted for publication in IEEE Software. This is the author's version which has not been fully edited and content may change prior to final publication. Citation information: DOI \href{https://doi.org/10.1109/MS.2026.3701516}{10.1109/MS.2026.3701516}.}%
}%
}%
}

\begin{document}

\sptitle{Theme Article: Special Issue on Engineering Agentic Systems}

\title{Agentic Code Generation: What We Hold and What We Miss}
\title{An Evaluation of Role-Based Multi-Agent Code Generation on Repository-Scale Problems}

\author{Benedetta Donato}
\affil{University of Milano-Bicocca, Milano, Italy}

\author{Noah Hagar-Dent}
\affil{The University of Auckland, Auckland, New Zealand}

\author{Aaron Worsnop}
\affil{The University of Auckland, Auckland, New Zealand}

\author{{Leonardo} Mariani}
\affil{University of Milano-Bicocca, Milano, Italy}

\author{Valerio Terragni}
\affil{The University of Auckland, Auckland, New Zealand}

\hypersetup{
pdftitle={An Evaluation of Role-Based Multi-Agent Code Generation on Repository-Scale Problems},
pdfauthor={Benedetta Donato; Noah Hagar-Dent; Aaron Worsnop; Leonardo Mariani; Valerio Terragni},
pdfsubject={This paper evaluates role-based multi-agent LLM code generation on repository-scale Java projects, comparing sequential and reflexive agentic workflows against a standalone LLM in terms of code size, compilation, similarity to developer implementations, scenario-intent representation, and resource use.},
pdfkeywords={agentic code generation; role-based multi-agent systems; repository-scale software engineering; Large Language Models; AI for software engineering; Java; program synthesis; code generation evaluation}
}

\markboth{Special Issue on Engineering Agentic Systems}{Special Issue on Engineering Agentic Systems}

\begin{abstract}\looseness-1 
Role-based multiagent code generation aims to make LLMs more effective on repository-scale problems, moving beyond small programming tasks. We evaluate this approach on 12 Java repositories, finding greater similarity to developer code than single LLMs, but a persistent gap from human implementations.
\end{abstract}

\maketitle

\chapteri{L}arge Language Models (LLMs) now support a wide range of software engineering activities, from requirements interpretation to code generation and maintenance~\cite{LLM4SEICSE2023}. Yet generating entire \emph{software systems} remains difficult: real-world repositories involve architectural structure, cross-file dependencies, build management, and non-functional constraints that amplify well-known LLM limitations: constrained context, hallucinations, brittle cross-file integration, and toolchain friction~\cite{LLMAgentsFSE2025,LLM4SEICSE2023}. LLMs also lack the high-level reasoning needed for complex requirements and large-scale architectural design~\cite{CodeDigitalTwin2025}, typically prioritising immediate functional specifications at the expense of more elaborate overarching requirements~\cite{LLMSUrveyACMTOSEM2024}, constraining the complexity of systems achievable without substantial human guidance~\cite{CodeDigitalTwin2025}.

A promising direction is \emph{agentic} code generation: distributing work across specialised LLM agents that mirror software team roles (e.g., requirements analyst, architect, developer, tester)~\cite{LLMAgentsCodeGenerationSurvey2025,LLMMASTOSEM2025}. 

A few recent frameworks have started exploring this direction: MetaGPT~\cite{MetaGPT}, AgileCoder~\cite{AgileCoder}, and RTADev~\cite{RTADev} explore multi-agent collaboration for code generation through structured workflows, Agile role assignments, and intention alignment, respectively. However, these frameworks have been evaluated on function-level benchmarks (e.g., HumanEval, MBPP \cite{HumanEval}) or small self-contained applications typically comprising fewer than 10 files and a few hundred lines of code. Successfully coordinating agents at repository scale remains an active research challenge, and there is still limited empirical understanding of when (and by how much) agentic approaches outperform standalone LLMs in realistic, repository-scale settings.

This paper investigates this tradeoff, comparing standalone LLMs to multi-agent architectures for large-scale code generation. 
To this end, we implemented a sequential and iterative multi-agent pipeline that decomposes requirements into implementable tasks, produces Java code, and integrates outputs into a buildable, compilable project. We then compared this agent team against an individual LLM on 12 real-world Java projects of varying sizes and complexities. 
The evaluation yielded insights into the current capabilities and limitations of LLM agents for large-scale code generation tasks. 

\section{Multi-Agent Code Generation}

\begin{figure*} [ht]
    \centering
    \includegraphics[width=\linewidth]{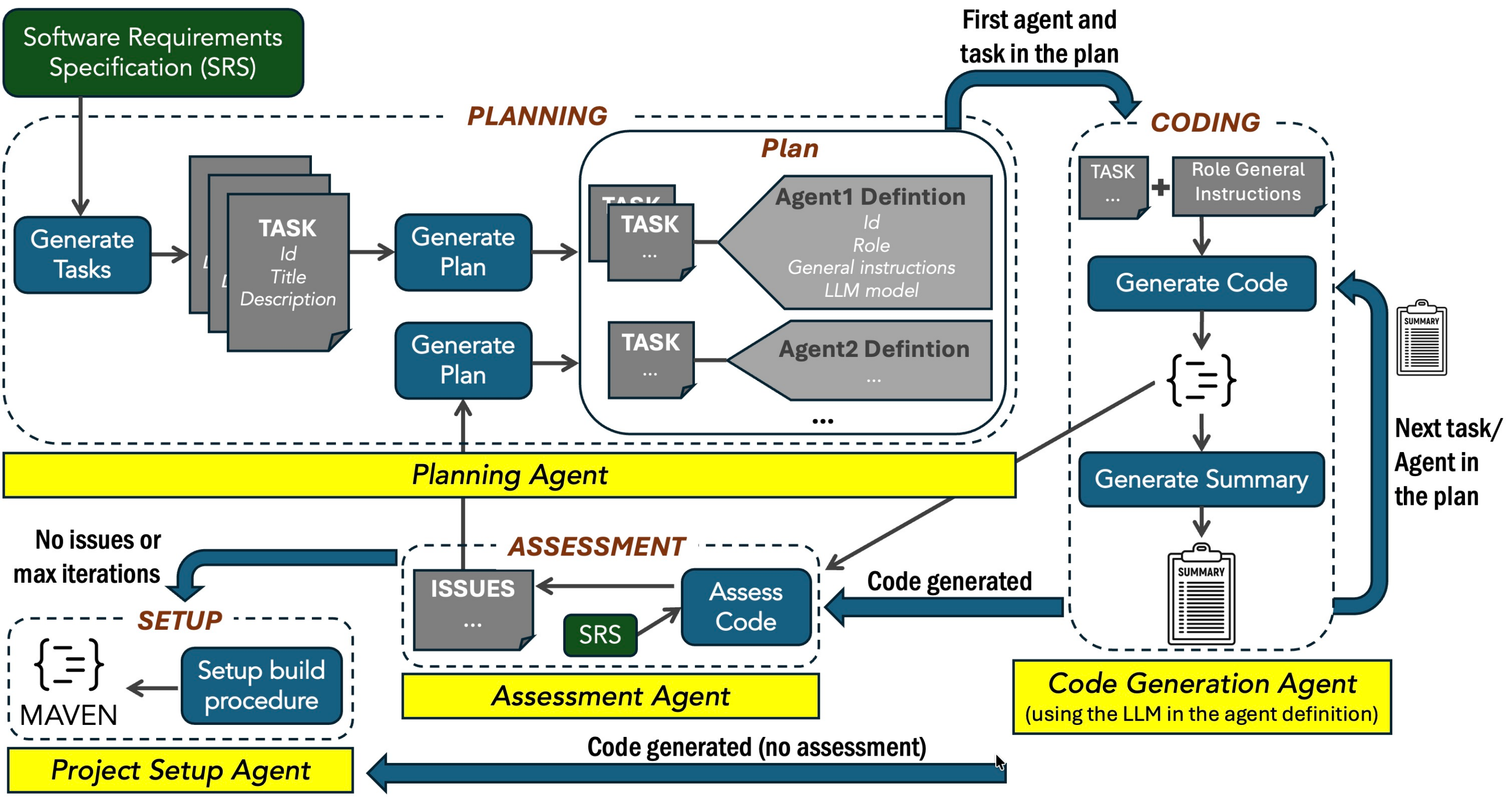}
    \caption{Logical architecture of Agentic Code Generation}
    \label{fig:overview}
\end{figure*}
\smallskip
Figure~\ref{fig:overview} illustrates the Agentic code generation process that we implemented. Similarly to other agentic code generation processes~\cite{selfplanning,Aflow},  it consists of four main phases: \emph{Planning}, which produces the plan that has to be completed to implement the application described in the input Software Requirements Specification (SRS); \emph{Coding}, which runs LLM agents according to the plan to obtain the implementation; \emph{Assessment}, which assesses the generated code, also against the input SRS; and \emph{Setup}, which generates the files necessary to build and run the generated code. We consider two variants of our flow, one \emph{Sequential}, where the assessment is skipped, and another \emph{Reflexive}, where the assessment phase is executed until no issues are found, or up to ten times. 
In all cases, the code generation starts from an SRS that describes the requirements as defined by the IEEE Standard on Software Requirements Specification~\cite{IEEEReq2018}: it includes an Introduction that describes the purpose, user base, project scope, and main features of the application, followed by the list of Functional and Non-Functional requirements.



\subsection{Planning}

In the planning phase, the planning agent is responsible for producing a plan from the input SRS, or from a list of detected issues when running after the assessment phase. When planning from the SRS, it first generates the set of tasks that must be completed to implement the SRS (\emph{Generate Tasks}). 

As a second step, the planning agent defines the code generation agents and assigns the identified tasks to them, with each agent executing one or more tasks (\emph{Generate Plan}). The tasks inside the plan are ordered. 

Each agent is defined by an identifier, a role, general instructions specific to its role, and the LLM to be used. 
The defined agent is then associated with some tasks based on its role. 
Roles are defined dynamically since the set of tasks to be completed to implement a requirement is not known in advance, and each task may require impersonating a different role.



When planning in response to code assessment, the list of produced tasks must solve the detected issues, rather than implementing the SRS.

\subsection{Coding}

For each task, the assigned LLM takes the agent's description and task details to generate the implementing code (\textit{Generate Code}), then summarizes the completed activity (\textit{Generate Summary}) and passes it to the next agent, ensuring awareness of existing code. From the second task onward, agents receive this summary as extra input. Prompts instruct agents to impersonate senior engineers and produce production-quality Java Maven code.

\subsection{Assessment}
\review{E.C3, E.C4, R1.C2}{In the assessment phase, the agent performs a static, text-based assessment of the generated code by comparing the implementation against the source SRS and reporting requirement-level issues, such as requirements that appear to be missing or only partially represented.} 
The list of all the detected issues is passed to the planning agent, which produces a plan to update the current implementation. 

\subsection{Setup}
This last phase finalizes code generation with the creation of a Maven project, including the generation of a \texttt{pom.xml} file. 


\section{Repository Selection and Design of the Experiment}

We first identified the top 1,000 Java GitHub repositories created in 2025, ranked by stars. This ensured that all projects postdate the training data cutoff of GPT-5 (September 30, 2024, according to OpenAI’s documentation\footnote{\url{https://platform.openai.com/docs/models/gpt-5}}), thereby reducing the risk of data leakage, i.e., the possibility that evaluated models had been exposed to the target implementations during training. Note that the data leakage issues cannot be entirely eliminated, since well-established frameworks (e.g., Spring, Maven, etc.) are extensively represented in the training data, and can also be present in the selected repositories. 
Table~\ref{tab:complexity_grouping} defines the criteria for each level: 
the Complex group captures roughly the top 5\% projects (3.8\% based on the thresholds); the Moderate group includes about 10\% of the projects (9.4\% based on the thresholds); the Simple group that includes around 20\% of projects (18.8\% based on the thresholds); and the remaining projects, which we label as Trivial. We then randomly selected three repositories per group, yielding a final set of 12 repositories\footnote{The twelve repositories and their identifiers are \textit{0ofo/java-memshell} (MemShell), \textit{Cooosin/AiClient} (AIClient), and \textit{IceC1eam/BUAAAutoSign} (AutoSign) as trivial repositories; \textit{CyrilFeng/karma} (Karma), \textit{GTyingzi/spring-ai-tutorial} (Spring), and \textit{uppnrise/distributed-rate-limiter} (Rate) as simple repositories; \textit{jd-opensource/JoySafety} (JoySafety), \textit{konmor/konmorReportServer} (KRS), and \textit{chaosblade-io/chaosblade-space-exploration} (Chaos) as moderate repositories; and \textit{a2aproject/a2a-java} (A2A), \textit{paohaijiao/jquick-pdf} (jquickpdf), and \textit{cbnbcbnb/RuoYi-Wvp} (RYW) as complex repositories}.

\medskip
To generate a structured SRS for each project, since SRS are rarely available in GitHub repositories, we used the README file and all Javadoc comments  as input to OpenAI GPT‑5. The list of the implemented functionalities, taken from the Javadoc, paired with high-level project descriptions, taken from the readme file, provide a fairly complete overview of the implemented requirements.  We manually validated each SRS to ensure proper structure: We checked that the content of each section is coherent with the section, and that no implementation details are disclosed in the SRS (e.g., interfaces, method signatures, classes, etc.). Although the SRSs cannot be as accurate as the ones written by developers, we could confirm content is sound, and no implementation details are leaked.

As a baseline, we consider how a standalone LLM performs on the same task, \emph{studying whether and how agentic solutions can improve the results obtained with LLMs}. The prompt used for individual LLMs is  the same used for agentic code generation with the only difference that the single LLMs are asked to generate code directly, while the agentic implementation first generates a plan.
To obtain comparable results, we used GPT‑5 (the most recent model at the time of the experiments) in its default configuration both as a standalone LLM and as the LLM powering the agentic approach. The reflexive version of agentic code generation is implemented using ChatDev 2.0\footnote{\url{https://github.com/OpenBMB/ChatDev}}.

\begin{table}[t]
\centering
\small
\setlength{\tabcolsep}{4pt}
\renewcommand{\arraystretch}{1.2}
\resizebox{\linewidth}{!}{
\begin{tabular}{l c c c c c}
\toprule
\textbf{Group} &
\textbf{LOC} &
\textbf{Class} &
\textbf{\makecell{Avg.\\ methods\\/class}} &
\textbf{\makecell{Avg.\\ CC}} &
\textbf{\makecell{\% methods\\ (CC $\geq$ 10)}} \\
\midrule
Trivial   & 0--2K    & 0--20     & 0--2     & 0--1     & 0\%--2\% \\
Simple    & 2K--10K  & 20--100   & 2--4     & 1--2.5   & 2\%--4\% \\
Moderate  & 10K--40K & 100--300  & 4--6     & 2.5--4   & 4\%--6\% \\
Complex   & $>40$K   & $>300$    & $>6$     & $>4$     & $>6\%$ \\
\bottomrule
\end{tabular}
}
\vspace{1.2mm}
\caption{Complexity group thresholds used for sampling.}
\label{tab:complexity_grouping}
\end{table}

To assess the results, we consider the following metrics: 


\emph{Process metrics:} we report the number of tasks and agents used in the code generation process;

\emph{Project size}: we report the number of generated LOCs, methods, and classes, including the percentage of classes that can be compiled immediately after the generation. 

\emph{Similarity}: There is no easy way of checking the equivalence between the generated code and the developers' implementation. 
For instance, any test available for the original program cannot be executed on the generated code, due to differences in the implemented entities, APIs, signatures, etc. that make tests impossible to execute. We thus relied on a combination of syntactic and semantic similarity metrics.
We computed code similarity using JPlag~\cite{JPLAG}, which is a plagiarism detection tool for Java, and CrystalBleu~\cite{CrystalBLEU}, which is a metric combining syntactic and data-flow similarities. Since we want to check the inclusion of the functionalities present in one project into the other, we used the JPlag max similarity which rewards inclusion as much as equality. For both metrics, the similarity is computed by identifying the most similar class for each generated class and then computing the average similarity across all classes. This strategy allows to match a same class to multiple classes, accounting for the cases where the same requirements are implemented with a different number of classes. 


\review{E.C1, R1.C1}{From a semantic perspective, we used an LLM-as-a-judge heuristic to estimate scenario-intent representation in the generated code. Specifically, we used Claude Code (with Sonnet 4.6) to analyze the original code and the generated implementations with respect to the input SRS. We instructed Claude Code to identify, for each requirement in the SRS, the scenarios represented in the code and to express them as Gherkin scenarios. Gherkin is the de-facto standard for behavior-driven development and provides a structured language for describing expected behavioral scenarios. We then used ChatGPT 5.4 to identify shared scenarios between the original and generated implementations.}

\review{E.C1, E.C2, R1.C1}{We report the rate of shared scenarios, defined as the ratio of scenarios identified in the original implementation that are also identified in the generated implementation. This metric should be interpreted as a measure of scenario-intent representation, not as a measure of functional correctness or executable behavior. In particular, a matched scenario indicates that the generated implementation textually or structurally appears to address a requirement-related behavior, but it does not necessarily imply that the behavior is correctly implemented.} 
\review{E.C1, R1.C1}{We inspected the output for a sample selection of projects to check that the metric reflects scenarios that are represented in both the original and generated code.}



\emph{Resources:} We consider the time, tokens, and actual cost consumed. Time is measured on a 2023 MacBook Pro (Apple M3, 16GB RAM). The models are executed remotely on OpenAI facilities.

Values are computed as an average over five repetitions, to account for the non-determinism of LLMs. 


\section{Code Generation Results}


\begin{table*}[ht]
\centering
\small
\setlength{\tabcolsep}{3pt}
\resizebox{\linewidth}{!}{
\begin{tabular}{ll | cc | ccc c| c c c | rcc}
\toprule
\textbf{Project} & \multirow{2}{*}{\textbf{Approach}} & \multicolumn{2}{|c|}{\textbf{Process}} & \multicolumn{4}{|c}{\textbf{Size}} & \multicolumn{3}{|c|}{\textbf{Similarity}} & \multicolumn{3}{c}{\textbf{Resources}} \\
                      \textbf{Group}       &                               & \textbf{\#Tasks} & \multicolumn{1}{r|}{\textbf{\#Agents}} & \textbf{LOC (wrt orig)}        & \textbf{\#Methods (wrt orig)} & \textbf{\#Classes (wrt orig)} & {\textbf{Compile}} & \textbf{JPlag Sim} & {\textbf{CrystalBleu}} & {\textbf{Scenario Intent}} & \textbf{Time} & \textbf{\#Tokens} & \textbf{Cost} \\ 
                          \midrule
\multirow{3}{*}{Trivial}  & \emph{LLM-Only}                          &    1         &    1                                &   1,448.1  (1.95$\times$)        &    69.7  (2.18$\times$)           &   13.3  (1.02$\times$) & {\textbf{34\%}}            &             0.08      &      { 0.01  }          &    {\textbf{94\%}}        & 17min&       38K   &       \$0.20                      \\
                          & \emph{Agentic-seq}                          & 33.9    & 5.7                          & 1,778.9 (2.39$\times$)  & 128.7 (4.02$\times$)     & 33.7 (2.59$\times$) & {24\%}       & 0.15 & {0.03 }&  {58\%}       & 52min         & 133K            & \$0.71             \\
                          &  {\emph{Agentic-refl}}                          &  {29.8}           &      {6 (8.2)}                              &   {\textbf{635.8 (0.85$\times$)}}         & {\textbf{59.3 (1.8$\times$)}}  & {\textbf{17.3 (1.3$\times$)}} &{33\%}&         {\textbf{0.31} }     &  {\textbf{0.10}}            & {86\%}                  & 42min &     104K                 &  \$0.56                          \\ \midrule
\multirow{3}{*}{Simple}   & \emph{LLM-Only}                             &     1         &     1                               &  \textbf{2,160.5 (0.15$\times$)}          &   117.7 (0.15$\times$)             &     31.1 (0.17$\times$)&{27\%}           & 0.10     & {0.08} &   {56\%}           &         22min       &         49K             &  \$0.26                          \\
                          & \emph{Agentic-seq}                            & 35.1             & 7.3                                   & 1,620.5 ($0.11\times$)          & \textbf{167.7 (0.21$\times$) }            & \textbf{37.1  (0.21$\times$)}&{19\%}              & 0.32  & {0.11} &   {58\%}             & 1h 08min                  & 189K                     & \$1.01                           \\
                          & {\emph{Agentic-refl} }                             &  {43.8  }          &  {8.2 (9.2)}                                  &   {1,068.7 (0.07$\times$) }       &  {98.0 (0.12$\times$) }  & {28.0 (0.16$\times$)} &{\textbf{43\%}}&   {\textbf{0.44}}      &   {\textbf{0.20}  }           & {\textbf{67\%}}                    &              1h 09min    &            156K          &  \$0.83                          \\  \midrule
\multirow{3}{*}{Moderate} &  \emph{LLM-Only}                         &        1      &   1                                 &  2,021.5 (0.06$\times$)      &       117.3 (0.04$\times$)         &     28.5 (0.11$\times$)  &{27\%}          & 0.32                  &   {0.12 }            &    {32\%}       & 21min &        46K  &     \$0.25                      \\
                          &  \emph{Agentic-seq}                               & 42.0             & 6.7                                   & \textbf{2,836.7 ($0.08\times$) }         & \textbf{243.8 (0.08$\times$) }              & \textbf{67.9  (0.25$\times$)}&{21\%}             & 0.40   & {0.17} &  {69\%}              & 1h 12min               & 211K                     & \$1.13                           \\
                          &  {\emph{Agentic-refl}  }                           &  {45.5 }          &     {7.2 (9.3)}                              &  {1,027.3 (0.03$\times$) }       &  {122.5 (0.04$\times$) }            &  {31.0 (0.11$\times$)} &{\textbf{40\%}}        &     {\textbf{0.45} }   & {\textbf{0.22}} &   {\textbf{85\%}}       &  1h 13min      &    159K                 &  \$0.85                         \\  \midrule
\multirow{3}{*}{Complex}  &  \emph{LLM-Only}                               & 1             & 1                                   &  2,236.8 (0.02$\times$)         &   132.5 (0.02$\times$)             &      27.1 (0.05$\times$) &{36\%}          &     0.43              &    {0.20 }           &   {61\%}       & 24min &      53K      &  \$0.28                          \\
                          &  \emph{Agentic-seq}                         & 53.7             & 7.7                                   & \textbf{3,064.3 ($0.03\times$)}          & \textbf{234.3   (0.04$\times$)}             & \textbf{62.1   (0.10$\times$)}&{25\%}             & 0.45 & {0.21}&  {34\%}             & 1h 48min               & 263K                     & \$1.40                           \\
                          &  {\emph{Agentic-refl}    }                           &  {51.7 }           &     { 6.6 (10)}                             &  {1,140.2 (0.01$\times$) }        & {125.2 (0.02$\times$)  }             & {38.0 (0.06$\times$)} &{\textbf{43\%}}               &{\textbf{0.50}}                   & {\textbf{0.25}}      & {\textbf{72\%}}&    1h 16min     &     197K                 &  \$1.05   
                          \\  \midrule 
\multirow{3}{*}{{Overall}} & {\emph{LLM-Only}}  & & & & & & {31\%} & {0.21} & {0.10} & {61\%}& & & \\
 &  {\emph{Agentic-seq}}  & & & & & & {22\%} & {0.33} & {0.13} & {55\%} & & & \\
 & {\emph{Agentic-refl}} & & & & & & {\textbf{40\%}} & {\textbf{0.43}} & {\textbf{0.19}} & {\textbf{78\%}}& & & \\

                          \bottomrule           
\end{tabular} %
}
\vspace{1.2mm}
\caption{\emph{Project Group} lists the project categories; \emph{Approach} specifies the code generation approach; \emph{\#Tasks} and \emph{\#Agents} indicate the average number of tasks and agents used for code generation (when applicable the number of executed iterations is reported between parentheses); \emph{Locs (wrt orig)}, \emph{Methods (wrt orig)}, \emph{Classes (wrt orig)} indicate the number of generated locs,  methods, and classes specifying within parenthesis the ratio with the locs, methods, and classes in the original repository; \emph{Compile} specifies the percentage of generated classes that compile; \emph{JPlag Sim}, \emph{CrystalBleu} report syntactic and data-flow similarity, \review{E.C1, R1.C1}{respectively; \textit{Scenario Intent} reports the scenarios matched by the LLM-as-a-judge heuristic and measures intent representation, not functional correctness}; 
\emph{Time}, \emph{Tokens} and \emph{Cost} report the time, tokens and approximated cost required by code generation.}
\label{tab:complexity_grouping}
\end{table*}

We compare agentic code generation (sequential or reflexive) to standalone LLMs. Table~\ref{tab:complexity_grouping} shows the results, considering metrics about the process
, the size of the generated code
, the similarity with the original developers' implementation
, and the resources used
. 
We report in bold the size values that are closers to the original implementation, and the best values for the other metrics.

\subsubsection{Generation Process} The number of tasks and agents scale with the size of the project, with Agentic-refl and Agentic-seq producing plans of similar sizes, ranging from 29.8 (for trivial projects) to 53.7 tasks (for complex projects). The number of used agents is similar, however agentic-seq uses agents multiple times during iterations, while agentic-seq uses agents only once.

We investigated if the summaries passed among the agents reflect the code that was generated. For each task, we measured the percentage of modified files that are nominated explicitly in the respective summary. The percentage ranged between 84.3\% for the Trivial projects to 91.2\% for Simpler projects, with Moderate and Complex projects scoring 86.7\% and 90.2\%, respectively. Note that this is a pessimistic estimate, since some changes could be described without referring to the name of the files that have been modified. Overall, this result confirms the suitability of summaries as an accurate mechanism for passing information among agents.

\subsubsection{Size of Generated Code}
Table~\ref{tab:complexity_grouping} column Size shows the locs, methods, and classes generated by the considered approaches. Agentic-refl tends to generate the smallest implementations, while Agentic-seq generates the largest ones, with LLM-Only producing implementation of intermediate sizes.


Manual inspection revealed that agentic approaches produce well-structured code with distinct logical layers (domain, service, control, support), whereas the standalone LLM tends to generate few large classes condensing multiple responsibilities.

With the growing size of the projects, the amount of generated code increases, although not in proportion to the expected project size. In fact, 
the generated code quantitatively represents a limited portion of the application (from 25\% to 1\% of the application). 

The delta in the amount of generated code shows that the strategy of planning followed by the execution of the planned tasks does not scale well to large projects, despite the presence of an iterative assessment and repair behavior. 

Code inspection showed that agentic approaches produce meaningful code capturing main domain abstractions and core features, but struggle with key architectural components (e.g., resolvers, adapters) and only partially implement features like error handling and edge cases. A significant portion of missing code corresponds to code that accumulated over time in the repositories (e.g., workarounds and partial refactorings).


\review{E.C3, E.C4, R1.C2}{Considering the percentage of classes that compile immediately after generation, Agentic-refl achieves the best result among the evaluated approaches. However, compilation success remains low overall. This is largely explained by the static nature of the reflection loop: since Agentic-refl does not use compiler diagnostics, stderr traces, stack traces, or execution feedback, it can address requirement-level omissions but cannot systematically detect and repair compilation errors such as missing imports, unresolved symbols, inconsistent method signatures, or API mismatches.}

In a nutshell, the obtained code can be used as a skeleton of the implementation that can be expanded by the developers, but not as a fully functioning system.

\subsubsection{Similarity to Original Code}
The results with both JPlag and CrystalBleu show that the agentic approaches generate implementations that are syntactically closer to the original code compared to the code generated by a standalone LLM. This is systematically confirmed for the projects in every group. Moreover, the reflexive implementation achieves higher similarity than the sequential implementation, giving evidence of how the corrective cycles can improve the quality of the code.

We investigated the per-project variance of the similarity, discovering a clear trend. 
The standard deviation of the JPlag similarity for Agentic-refl consistently scored 0.01, while it ranged between 0.04 and 0.06 for LLM-Only, and between 0.06 and 0.1 for Agentic-seq. This shows how the refinement loop tends to make the code generation task converges towards consistent implementations, although the initial agentic implementation of a SRS may vary a lot. LLM-only produces relatively stable results, probably benefiting of direct code generation, not having to go through a plan that introduces variability and synchornization issues.

\review{E.C1, R1.C1}{The scenario-intent metric shows that the LLM-only approach textually represents a good proportion of the scenarios identified in the original implementation, especially for trivial projects.}
\review{E.C1, R1.C1}{Agentic-seq instead struggles to represent the same scenario intents in the generated code.}
While Agentic-refl performs the best in all the cases, with the exception of trivial programs.

Iterative code improvement is thus essential to obtain better code, especially when projects are large and complex, as witnessed by an increasingly bigger delta between Agentic-refl and both LLM-Only and Agentic-seq.  


\review{E.C1, E.C2, R1.C1, R1.C2}{Considering size, similarity, and compilation metrics together, agentic workflows tend to produce better-structured code and to represent more scenario intents than the standalone LLM. However, these scenario-intent matches should not be interpreted as evidence of executable behavior: the generated implementations are often partial and require human intervention, especially because a substantial fraction of the generated classes does not compile immediately.}




\subsubsection{Resources}

Table~\ref{tab:complexity_grouping} shows the resources (time, tokens, and cost) spent by the agents in the code generation tasks.

Agentic code generation is relatively fast, requiring between 52mins and 1h 48min, depeding on the size of the project. The number of tokens consumed per project varies proportionally with time, with a cost ranging from \$0.20 for trivial projects to \$1.40 for complex ones. 
 


Compared with a standalone LLM, which is fast and requires few tokens, an agentic approach is naturally more resource‑intensive. 
Nonetheless, the additional overhead is acceptable given the improved similarity to the developers’ ground‑truth implementations. 

Sometimes, the tokens and time consumed by the sequential agentic workflow are slightly higher than those consumed by the reflexive workflow. This is an example of how specific platforms may induce changes in the behavior of the agents, due to the way agentic roles (e.g., planner, programmer, and reviewer) are defined in the system prompts.

Resource-wise, an aspect playing a relevant role is the human effort necessary to complete an implementation. Although precisely assessing this component is not in the scope of this study, the scenarios metric is informative. 
\review{E.C1, R1.C1}{In fact, Agentic-refl represented 78\% of the scenario intents identified by the LLM-as-a-judge heuristic.} 
\review{E.C1, E.C2, R1.C1, R1.C2}{That is, 22\% of the scenario-intents are not represented in the generated code, while the remaining ones correspond to partial code fragments that may provide a starting point for developers, but not necessarily executable or functionally correct behavior.} Based on our inspection, the generated code tends to cover certain aspects very superficially. For instance, the domain classes are often simpler than the ones in the original implementation, and cross-cutting concerns, such as the implementation of the authorization and security layers, are drastically simplified. This demands the developers' intervention, either by interacting with the agentic systems or manually implementing the missing functionalities.

We have made all experimental data and implementations publicly available to support future research in this field\footnote{\url{https://drive.google.com/drive/folders/1GL2BksvB7OH-z0k_SF3YC3bI6Z88AV_0?usp=sharing} 
.}.

\section{Lessons Learned}
Our results yield actionable lessons learned relevant to both practitioners and researchers. 

\textbf{Agentic AI with reflexive capabilities outperforms standalone LLMs:} 
LLM-Only performed well for Trivial and Simple projects, but 
\review{E.C1, R1.C1}{Agentic systems that integrate the reflexive component performed better in terms of structure and scenario-intent representation for Medium and Complex systems.}
Our inspection revealed that the code generated by the agentic approach is better designed and structured, compared to the code generated by a standalone LLM, which tends to be monolithic. 

\textbf{Agentic AI can over-engineer simple applications:} If a demo application or a simple proof-of-concept is needed, the agentic approach may over-engineer the code, introducing unnecessary features (e.g., about telemetry and security), or over-complicating the design, introducing an unnecessary number of layers and components. Conversely, the inclusion of these features by default is a virtue when considering professional code development.

\textbf{Agentic AI is yet unable to automatically generate an actually working system:} Agentic code generation performed better than standalone LLMs, but the obtained implementations were partial, especially for larger repositories. 
\review{E.C2, E.C3, E.C4, R1.C2}{Although Agentic-refl improved scenario-intent representation and achieved the highest compilation rate among the evaluated approaches, compilation success remained low overall. This limitation is mainly due to the static nature of the current reflection loop, which does not use compiler feedback and dynamic (e.g., test execution) data.} 

\textbf{Agentic AI can be leveraged to obtain an initial (partial) implementation of an application:} The Agentic approach can quickly generate an implementation that captures the key components, abstractions, and domain elements of a system. This can be a good basis to start the development from. 


\textbf{Agents can complete tasks efficiently:} The resources (time, tokens, and cost) needed to generate the code that implements the SRS were quite limited. It was possible to obtain an initial implementation of the project in less than 2 hours in the majority of cases, spending less than 2 dollars per project. Yet, developers have to consider spending significant effort to finalize the implementation of the project.

\textbf{Human-in-the-loop scenarios are still the most practical choice:} The possibility of generating a complete system from a brief SRS document still requires substantial research. For instance, detailed SRSs can potentially help obtaining better systems. Providing implementation‑specific prompts, such as guidance on relevant edge cases, required architectural elements, and cross‑cutting concerns (e.g., logging, security, etc.), can substantially improve the quality of the generated systems. Iterative human–AI interaction can also be leveraged to progressively refine and complete partial implementations. Given the partial systems produced in our experiments, these strategies currently remain the most promising options to investigate for cost‑effectively obtaining working systems.


\smallskip

In conclusion, this study highlights fully-automated repository-scale agentic code generation as a promising direction for the future. 
\review{E.C5, R1.C2}{However, additional studies and more work are needed to thoroughly investigate agentic code generation and to define execution-aware generation processes, such as compiler- and test-in-the-loop reflection workflows, that can deliver a working system from an SRS document.}

\def\refname{REFERENCES}

\vspace*{-8pt}


\begin{IEEEbiography}{\,}
\setlength{\columnsep}{0.08in}
\begin{wrapfigure}[9]{l}{1in}
\vspace{0.em}
\includegraphics[width=1in,height=1.25in, keepaspectratio]{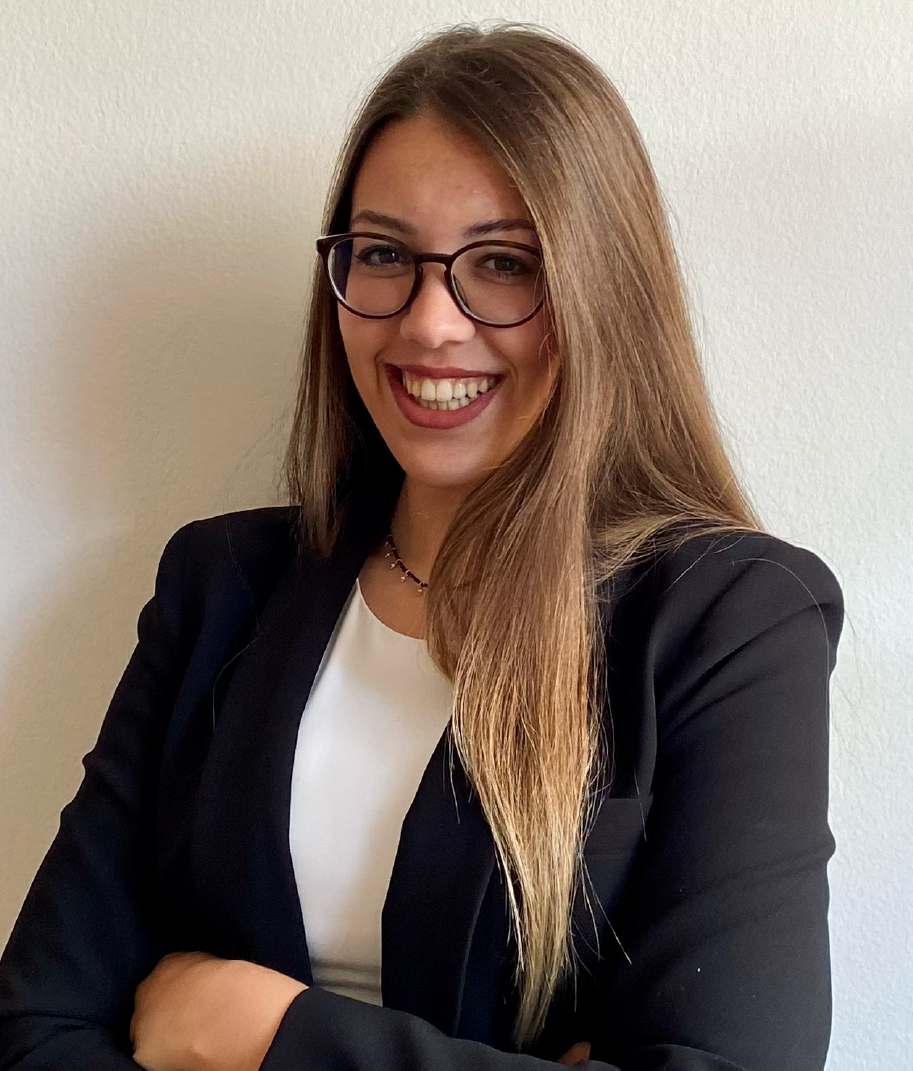}
\vspace{-1em}
\end{wrapfigure}

\noindent\textbf{Benedetta Donato} {\,} is a Ph.D. student in Computer Science at the University of Milano-Bicocca. Her current research interests include AI-assisted software engineering, Large Language Models for code generation, and empirical studies on developer--AI collaboration. She received the M.Sc. degree in Computer Science from the University of Milano-Bicocca in 2024, with a thesis focused on the impact of generation parameters on the quality and correctness of code produced by LLMs. She is an IEEE Student Member. Contact information: benedetta.donato@unimib.it.
\end{IEEEbiography}

\vspace{6mm}

\begin{IEEEbiography}{\,}
\setlength{\columnsep}{0.08in}
\begin{wrapfigure}[8]{l}{1in}
\vspace{1.1em}
\includegraphics[width=1in,height=1.25in, keepaspectratio]{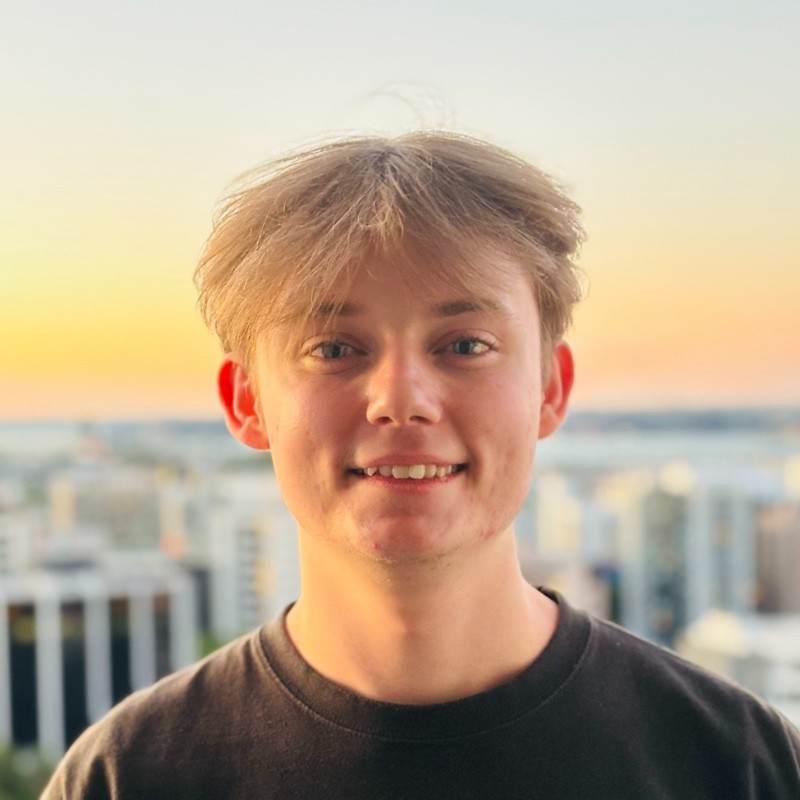}
\vspace{-1em}
\end{wrapfigure}

\noindent\textbf{Noah Hagar-Dent} {\,}
is a recent graduate of the University of Auckland, New Zealand. His research interests include agentic software engineering. He holds a Bachelor of Engineering (Honours) degree in Software Engineering, with a research project focusing on agentic AI, awarded by the University of Auckland. He can be contacted at nhag609@aucklanduni.ac.nz.
\end{IEEEbiography}

\begin{IEEEbiography}{\,}
\setlength{\columnsep}{0.08in}
\begin{wrapfigure}[8]{l}{1in}
\vspace{0.em}
\includegraphics[width=1in,height=1.25in, keepaspectratio]{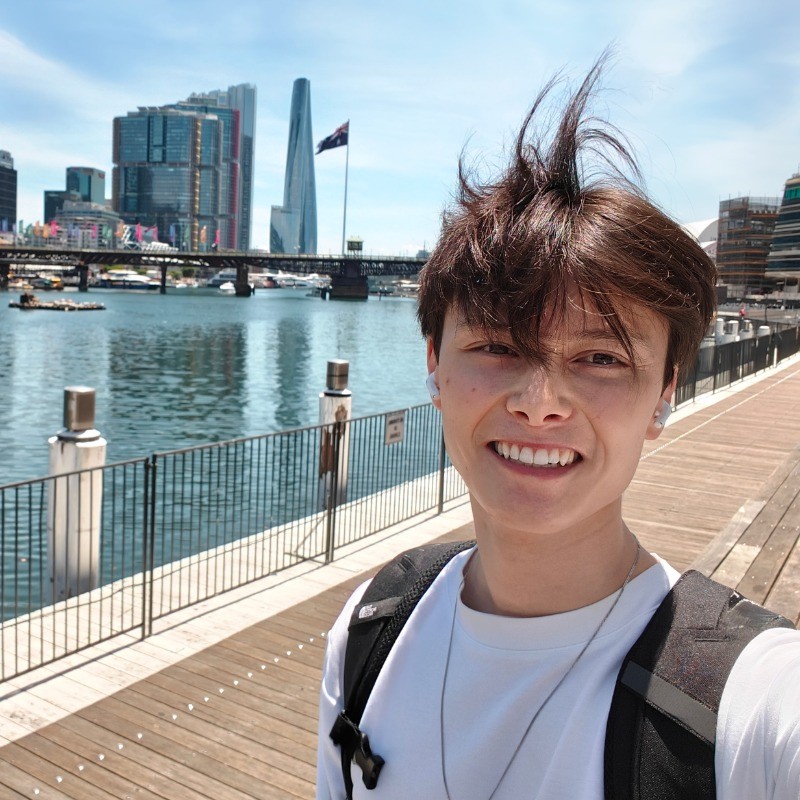}
\vspace{-1em}
\end{wrapfigure}

\noindent\textbf{Aaron Worsnop} {\,} is a recent graduate of the University of Auckland, New Zealand.. His research interests include agentic software engineering. He holds a Bachelor of Engineering (Honours) degree in Software Engineering, with a research project focusing on agentic AI, awarded by the University of Auckland. He can be contacted at awor776@aucklanduni.ac.nz.
\end{IEEEbiography}

\begin{IEEEbiography}{\,}
\setlength{\columnsep}{0.08in}
\begin{wrapfigure}[7]{l}{1in}
\vspace{0.em}
\includegraphics[width=1in,height=1.25in, keepaspectratio]{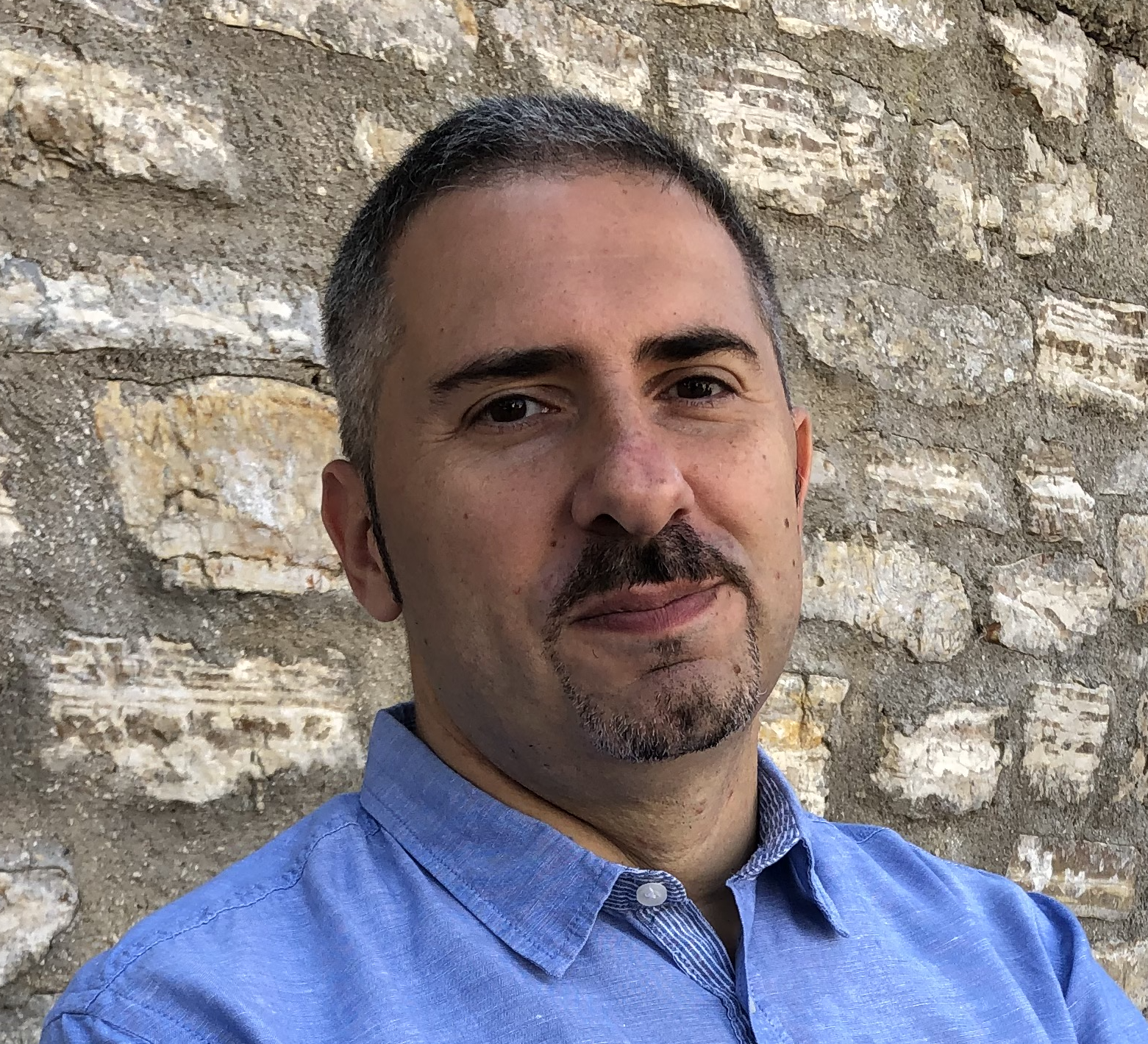}
\vspace{-1em}
\end{wrapfigure}

\noindent\textbf{Leonardo Mariani} {\,} is Full Professor at the University of Milano - Bicocca. He obtained his Ph.D. in Computer Science from the same university. His main research interests concern the interplay between AI and software engineering, test case generation for conversational systems, and quality assurance of mobile applications. He is an IEEE Senior Member. Contact information: leonardo.mariani@unimib.it.
\end{IEEEbiography}

\begin{IEEEbiography}{\,}
\setlength{\columnsep}{0.08in}
\begin{wrapfigure}[8]{l}{1in}
\vspace{0.em}
\includegraphics[width=1in,height=1.25in, keepaspectratio]{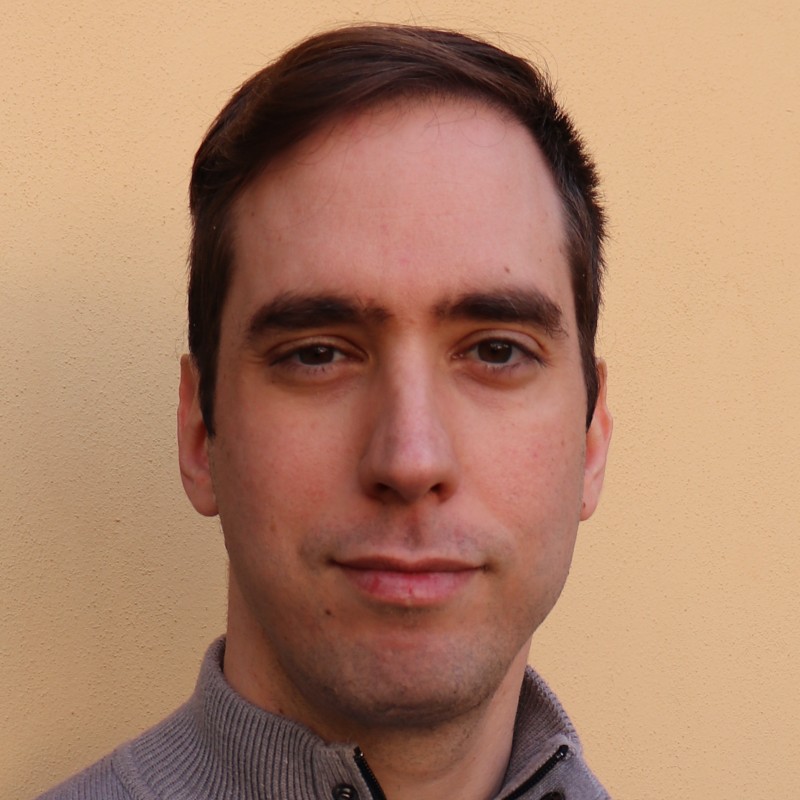}
\vspace{-1em}
\end{wrapfigure}

\noindent\textbf{Valerio Terragni}{\,}
is a Senior Lecturer at the University of Auckland, New Zealand. His main research interests concern AI for software engineering, test case generation, and metamorphic testing. He obtained his Ph.D. in Computer Science from The Hong Kong University of Science and Technology, with a dissertation focusing on automated software testing. He is a member of the IEEE Computer Society and the ACM. Contact information: v.terragni@auckland.ac.nz.
\end{IEEEbiography}

\end{document}